\documentclass{article}
\usepackage{amssymb}
\usepackage{amsmath}
\numberwithin{equation}{section} 
 \DeclareMathOperator{\sgn}{\rm sgn}
 \DeclareMathOperator{\diag}{\rm diag}

 \DeclareMathOperator{\Rs}{\mathbb{R}}
 \DeclareMathOperator{\Cs}{\mathbb{C}}

 \DeclareMathOperator{\Zs}{\mathbb{Z}}

 \DeclareMathOperator{\Go}{\mathcal{G}}

\def\t#1{\widetilde{#1}}
\def\h#1{\widehat{#1}}
\def\ol#1{\overline{#1}}

\title{Hirota difference equation: IST, Darboux transformation and solitons}
\author{A.~K.~Pogrebkov \\Steklov Mathematical Institute, Moscow, and\\National Research University Higher School of Economics,\\ International Laboratory of Representation Theory and\\ Mathematical Physics, \\20 Myasnitskaya Ulitsa, Moscow 101000, Russia\\ pogreb@mi.ras.ru}
\date{\today}
\begin{document}
\maketitle 
\begin{abstract}
Direct and inverse problems for the Hirota difference equation are considered. Jost solutions and scattering data are introduced and their properties are presented. Darboux transformation in a special case  is shown to give evolution with respect to discrete time and a recursion procedure for consequent construction of the Jost solution at arbitrary time, if the initial value is given. Some properties of the soliton solutions are discussed.
\end{abstract}


\section{Introduction}

Hirota bilinear difference equation (HBDE) was introduced as equation on the $\tau$-function in~\cite{Hirota1,Hirota2}. In a slightly different notation we write it here in the form:
\begin{equation}
\tau^{(1)}(m)\tau^{(2,3)}(m)+\tau^{(2)}(m)\tau^{(3,1)}(m)+\tau^{(3)}(m)\tau^{(1,2)}(m)=0,\label{Ht}
\end{equation} 
where $\tau(m)=\tau(m_1,m_2,m_3)$ is a function of 3 numbers (independent variables) $m_1,m_2,m_3\in\Zs$. Here and below upper indexes 1,2, и 3 in parenthesis denote unity shift
\begin{equation}
\tau^{(1)}(m)=\tau(m_1+1,m_2,m_3),\qquad \tau^{(2,3)}(m)=\tau(m_1,m_2+1,m_3+1)\ldots,\label{upper} 
\end{equation}
of the variable with the corresponding number. After works~\cite{Hirota2,Miwa} this equation is known to generate many discrete and continuous integrable equations, such as KP, mKP, two-dimensional Toda lattice, sine-Gordon, Benjamin--Ono, etc.,  by means of special limiting procedures. Because of this HBDE is often considered to be a fundamental integrable system. In~\cite{bk} Hirota bilinear difference equation is called the ``generalized KP hierarchy.'' This equation also appears as the model-independent functional relation for eigenvalues of quantum transfer matrices. Detailed survey of the results refered to this equation is given in~\cite{zabrodin1,zabrodin2}, see also citations therein. Octahedral structure of HBDE is studied in \cite{Saito}. Its elliptic solutions were considered in~\cite{kwz}.

Following \cite{zabrodin2} we introduce function $v(m)=v(m_1,m_2,m_3)$ by means of equalities
\begin{align}
&v_1(m)-v_3(m)=\dfrac{\tau^{(1,3)}(m)\tau(m)}{\tau^{(1)}(m)\tau^{(3)}(m)},\label{vt1} \\
&v_2(m)-v_1(m)=\dfrac{\tau^{(1,2)}(m)\tau(m)}{\tau^{(2)}(m)\tau^{(1)}(m)}.\label{vt2} 
\end{align}
Here and below besides (\ref{upper}) we use notation of the kind
\begin{align}
&v_1(m)=v(m_1+1,m_2,m_3)-v(m)\equiv v^{(1)}(m)-v(m),\nonumber\\ 
&v_2(m)=v(m_1,m_2+1,m_3)-v(m)\equiv v^{(2)}(m)-v(m),\ldots\label{fd} 
\end{align}
for the first finite differences. It is clear that (\ref{vt2}) follows from (\ref{vt1}) by cyclic permutation of indexes $\{1,2,3\}$. Then  the third equation
\begin{equation}
v_3(m)-v_2(m)=\dfrac{\tau^{(2,3)}(m)\tau(m)}{\tau^{(2)}(m)\tau^{(3)}(m)},\label{vt3} 
\end{equation}
obtained by cyclic permutation is equivalent to (\ref{Ht}) thanks to (\ref{vt1}) and (\ref{vt2}) . Then it is easy to check that function $v(m)$ obeys the following nonlinear equation:
\begin{equation}
 (v_{1}-v_{2})v_{1,2}+(v_2-v_3)v_{2,3}+(v_3-v_1)v_{3,1}=0,\label{Hv} 
\end{equation} 
where notation for the second differences of the kind 
\begin{equation}
v_{1,2}(m)=(v_1)_2(m)=(v_2)_1(m)\label{lr}
\end{equation}
was introduced. Relations (\ref{vt1})--(\ref{vt3}) are known to admit gauge invariance:
\begin{equation}
\tau(m)\to f_1(m_1)f_2(m_2)f_3(m_3)\tau(m).\label{gauge} 
\end{equation} 

Eq. (\ref{Hv}) has Lax representation (see, say, \cite{zabrodin1}) with the Lax pair which is given by any two of the following three equations:
\begin{align}
 \varphi_2(m,k)&=\varphi_1(m,k)+\bigl(v_2(m)-v_1(m)\bigr)\varphi(m,k),\label{Lv1} \\
 \varphi_3(m,k)&=\varphi_1(m,k)+\bigl(v_3(m)-v_1(m)\bigr)\varphi(m,k),\label{Lv2}\\ 
 \varphi_3(m,k)&=\varphi_2(m,k)+\bigl(v_3(m)-v_2(m)\bigr)\varphi(m,k),\label{Lv3} 
\end{align}
where $k\in\Cs$ is a spectral parameter. Eq. (\ref{Hv}) can be considered as an evolution equation, where, say, $m_1$ and $m_2$ play the role of space variables, and $m_3$ is the time one. It is reasonable to consider the Cauchy problem for (\ref{Hv}) with initial data
\begin{equation}
v(m_1,m_2,0)=v_0(m_1,m_2).\label{CP} 
\end{equation}  
But in this formulation this problem is ill posed. Indeed, it is easy to see that it has two trivial solutions: $v(m)=v_0(m_1+m_3,m_2)$ and $v(m)=v_0(m_1,m_2+m_3)$. On the other side $\tau$-function representation (\ref{vt1}) does not allow solutions with $v_i=v_j$, $i\neq{j}$. Thus equations (\ref{Ht}) and (\ref{Hv}) are not equivalent and some additional conditions on the class of solution of this Cauchy problem must be imposed. It is reasonable to mention that the evolution form (\ref{Hv}) of the Hirota difference equation is also bilinear, as HBDE (\ref{Ht}). 

In \cite{commut3} the Hirota difference equation appeared as a result of the dressing of a some commutator identity.  Let an associative algebra with unity $I$ over $\Cs$ is given. Choose any element $A$ of this algebra such that in this algebra there exist $(A-a_1I)^{-1}$, $ (A-a_2I)^{-1}$, and $ (A-a_3I)^{-1}$, where $a_1$, $a_2$, and $a_3$, are some complex parameters.  For any , $B$ we introduce operations (commutators in the group sense)
\begin{equation}
  \delta_{a_j}(B)=(A-a_{j})B(A-a_j)^{-1}-B\label{2}
\end{equation}
 on the algebra (here and below we omit the unity multipliers of $a_j$). Product of operations (\ref{2}) will be understood as composition: $(\delta_{a_1}\delta_{a_2})(B)\equiv\delta_{a_1}(\delta_{a_2}(B))$, etc.  Then as a trivial consequence of associativity we get identity
\begin{equation}
  (a^{}_1-a^{}_2)\delta_{a_1}\delta_{a_2}+(a^{}_2-a^{}_3)\delta_{a_2}\delta_{a_3}+(a^{}_3-a^{}_1)
  \delta_{a_3}\delta_{a_1}=0,\label{3}
\end{equation}
 
Now, taking commutativity of operations (\ref{2}) into account we introduce dependence on discrete ``times'' $m_n\in\Zs$, $n=1,2,3$ by means of the equalities:
\begin{equation}
B(m_1,m_2,m_3)=\Biggl(\prod_{n=1}^{3}(A-a_n)^{m_n}\Biggr)B\Biggl(\prod_{n=1}^{3}(A-a_n)^{m_n}\Biggr)^{-1}.
\label{4}
\end{equation}
Because of (\ref{3}) this function of three variables obeys the linear difference equation
\begin{equation}
(a^{}_1-a^{}_2)B_{1,2}(m)+(a^{}_2-a^{}_3)B_{2,3}(m)+(a^{}_3-a^{}_1)B_{1,3}(m)=0.\label{lB} 
\end{equation} 
Following the method of extended resolvent (see \cite{first}--\cite{KP-JMP}) and \cite{commut1,commut2} we introduced in \cite{commut3} a kind of a dressing procedure for Eq.~(\ref{lB}) that lead us to the nonlinear equation
\begin{equation}
\bigl[a_1-a_2-u_{1}+u_{2}\bigr]u_{1,2}+\bigl[a_2-a_3-u_2+u_3\bigr]u_{2,3}+\bigl[a_3-a_1-u_3+u_1\bigr]u_{3,1}=0\label{Hd} 
\end{equation}
on a function $u(m_{1},m_{2},m_{3})$ of three discrete variables that decay rapidly enough with growing of any $m_i$. This equation is a ``nonlinearization'' of (\ref{lB}) and below we consider the Cauchy problem 
\begin{equation}
u(m_1,m_2,0)=u_0(m_1,m_2),\label{CPu} 
\end{equation}
for real initial data $u_0$ decaying rapidly enough. 

It is clear that in the case $a_1=a_2=a_3$ Eq.~(\ref{lB}) becomes senseless. On the other side in this case Eqs. (\ref{Hv}) and (\ref{Hd}) coincide and we meet with the ill-posedness  of the Cauchy problem discussed above. Moreover, if, say, $a_3=a_1\neq{a_2}$ we have two obvious solutions of the Cauchy problem (\ref{Hd}) and (\ref{CPu}): $u(m)=u_{0}(m_1+m_3,m_2)$ and $u(m)=u_{0}(m_1,m_2+m_3)+(a_1-a_2)m_3$, so again this Cauchy problem is ill posed. In order to exclude such cases we impose further on condition
\begin{equation}
a_{i}\neq a_{j},\quad i\neq j,\quad i,j=1,2,3.\label{cond} 
\end{equation} 
By means of the Inverse scattering transform (IST) we show below that this condition is also sufficient for solvability of the Cauchy problem in the class of rapidly decaying $u(m)$. Taking that  Eq. (\ref{Hd}) reduces to (\ref{Hv}) by means of substitution
\begin{equation}
v(m)=u(m)-a_1m_1-a_2m_2-a_3m_3\label{vu} 
\end{equation}
into account, we see that $v(m)$ is linearly growing and
\begin{equation}
v_j(m)\to-a_j,\quad m\to\infty,\quad j=1,2,3.\label{vinf} 
\end{equation}
Thus condition (\ref{cond}) excludes possibility of any of equalities $v_i=v_j$ for any $i\neq{j}$. Thanks to (\ref{vu}) this substitution we below call both eqs.~(\ref{Hv}) and (\ref{Hd}) the Hirota difference equation. We consider here real solutions of this equation, so in what follows we impose condition of reality on all constants $a_j$  (see \cite{commut3}):
\begin{equation}
a_1,a_2,a_3\in\Rs.\label{condr} 
\end{equation} 

Thanks to (\ref{vu}) relations (\ref{vt1})--(\ref{vt3})  are written in the form
\begin{align}
&u_1-u_3=a_1-a_3+\dfrac{\tau^{(1,3)}\tau}{\tau^{(1)}\tau^{(3)}},\label{ut1} \\
&u_2-u_1=a_2-a_1+\dfrac{\tau^{(1,2)}\tau}{\tau^{(2)}\tau^{(1)}},\label{ut2} \\
&u_3-u_2=a_3-a_2+\dfrac{\tau^{(2,3)}\tau}{\tau^{(2)}\tau^{(3)}},\label{ut3} 
\end{align}
and under the same substitution equations (\ref{Lv1})--(\ref{Lv3}) are
\begin{align}
 \varphi_2(m,k)&=\varphi_1(m,k)+\bigl(u_2(m)-u_1(m)+a_1-a_2\bigr)\varphi(m,k),\label{55} \\
 \varphi_3(m,k)&=\varphi_1(m,k)+\bigl(u_3(m)-u_1(m)+a_1-a_3\bigr)\varphi(m,k),\label{56} \\
\varphi_3(m,k)&=\varphi_2(m,k)+\bigl(u_3(m)-u_2(m)+a_2-a_3\bigr)\varphi(m,k).\label{55'} 
\end{align}
We choose two first equations as the Lax pair for the Hirota difference equation (\ref{Hd}).

The article is organized as follows. In Sec.~2 we introduce Jost solutions and study their properties and properties of the Green's function (Secs.~2.1 and 2.2). In Sec. 2.3 we introduce scattering data and present their properties, including time evolution. This enables us to formulate Inverse problem there and to introduce the generating functional of the integrals of motion. In Sec.~3 we consider Darboux transformation of the continuous specter and show that for the Hirota equation this transformation gives, in particular, time evolution. Moreover, it can be considered as recursion procedure for step-by-step construction of the Jost solution for an arbitrary time, if the initial value is known. In Sec.~3.2 we present relations of the IST objects with the $\tau$-function approach. In Sec.~4 we consider properties of the soliton solutions for this equation and discuss some opened problems in description of these solutions.

\section{Direct and inverse problems}

\subsection{Green's function and Jost solution}
Jost solution $\varphi(m,k)$ of (\ref{55}) reads as
\begin{align}
&\varphi(m,k)=E(m,k)\chi(m,k),\qquad k\in\Cs,\label{58:1}\\
\intertext{where}
&E(m,k)=(k-a_1)^{m_1}(k-a_2)^{m_2}(k-a_3)^{m_3},\label{58:2}\\
\intertext{and is fixed by a condition that function $\chi(m,k)$ obeys normalization}
&\lim_{k\to\infty}\chi(m,k)=1.\label{58:3} 
\end{align}
In terms of the function $\chi(m,k)$ equations (\ref{55}) and (\ref{56}) of the Lax pair take the form
\begin{align}
 (k-a_2)\chi_2(m,k)&=(k-a_1)\chi_1(m,k)+\bigl(u_2(m)-u_1(m)\bigr)\chi(m,k),\label{58:5} \\
 (k-a_3)\chi_3(m,k)&=(k-a_1)\chi_1(m,k)+\bigl(u_3(m)-u_1(m)\bigr)\chi(m,k),\label{58:6}
\intertext{and (\ref{55'}) as}
(k-a_3)\chi_3(m,k)&=(k-a_2)\chi_2(m,k)+\bigl(u_3(m)-u_2(m)\bigr)\chi(m,k),\label{60:1}
\end{align}
preserving invariance with respect to the cycle permutations of the indexes $\{1,2,3\}$. 

The ``integral'' equation on $\chi(m,k)$ that determines solution of (\ref{58:5}) obeying (\ref{58:3}) was derived in~\cite{commut2}. Here we write it in the following form
\begin{align}
\chi(m,k)&=1+\sum_{n_1,n_2\in\Zs}G(m-n,k)(u_2(n)-u_1(n))\chi(n,k),\quad k\in\Cs,\label{di3:1}
\intertext{where the Green's function is equal to} 
G(m,k)&=\oint\limits_{|\zeta_1|=1}\dfrac{d\zeta_1}{2\pi i}\oint\limits_{|\zeta_2|=1}\dfrac{d\zeta_2}{2\pi i}\,
\dfrac{\zeta^{m_1-1}_1\zeta^{m_2-1}_2}{(k-a_2)\zeta_2-(k-a_1)\zeta_1+a_2-a_1}.\label{di4} 
\end{align}
Denominator of the integral in the r.h.s.\ of this equality has zeros in the two cases only:
\begin{equation}
\zeta_1=\zeta_{2}=1,\quad\text{or}\quad \zeta_1=\dfrac{\ol{k}-a_1}{k-a_1},\quad\zeta_2=\dfrac{\ol{k}-a_2}{k-a_2},\label{di4'} 
\end{equation} 
so the integral converges and defines $G(m,k)$ as distribution of $k$. Integrating by $\zeta_1$ or $\zeta_2$ we get two representations
\begin{align}
G(m,k)&=\dfrac{1}{k-a_1}\oint\limits_{|\xi|=1}\dfrac{d\xi}{2\pi i}\Bigl(\dfrac{(k-a_2)\xi+a_2-a_1}{k-a_1}\Bigr)^{m_1-1}
\xi^{m_2-1}\times\nonumber\\
&\quad\times\bigl[\theta(m_1\leq0)\theta\bigl(|(k-a_2)\xi+a_2-a_1|-|k-a_1|\bigr)-\nonumber\\
&\quad-\theta(m_1\geq1)\theta\bigl(|k-a_1|-|(k-a_2)\xi+a_2-a_1|\bigr)\bigr]\equiv\label{gr1}\\
&\equiv\dfrac{1}{k-a_2}\oint\limits_{|\xi|=1}\dfrac{d\xi}{2\pi i}\xi^{m_1-1}
\Bigl(\dfrac{(k-a_1)\xi+a_1-a_2}{k-a_2}\Bigr)^{m_2-1}\times\nonumber\\
&\quad\times\bigl[-\theta(m_2\leq0)\theta\bigl(|(k-a_1)\xi+a_1-a_2|-|k-a_2|\bigr)+\nonumber\\
&\quad+\theta(m_2\geq1)\theta\bigl(|k-a_2|-|(k-a_1)\xi+a_1-a_2|\bigr)\bigr],\label{gr2} 
\end{align}
where $\theta$ of continuous argument denotes the standard Heaviside step function and for the discrete variables it is the characteristic function of the corresponding interval:
\begin{equation}
\theta(m_1\leq0)=\left\{\begin{array}{ll}
0,& m_1\geq1,\\
1,& m_1\leq0,
\end{array}\right.,\qquad 
\theta(m_1\geq1)=\left\{\begin{array}{ll}
1,& m_1\geq1,\\
0,& m_1\leq0,
\end{array}\right.,\qquad\text{etc.}\label{gr3} 
\end{equation} 
Any of these representations shows that thanks to (\ref{condr}) the Green's function has properties of  conjugation
\begin{equation}
\ol{G(m,k)}=G(m,\ol{k})=\Bigl(\dfrac{k-a_1}{\ol{k}-a_1}\Bigr)^{m_1}\Bigl(\dfrac{k-a_2}{\ol{k}-a_2}\Bigr)^{m_2}G(m,k)\label{gr4}
\end{equation} 
and antisymmetry 
\begin{equation}
G(m_1,m_2,k)=-G(m_2,m_1,k)\Bigr|_{a_1\leftrightarrow{a_2}}.\label{gr5}
\end{equation} 

Asymptotic behavior of the Green's function follows from (\ref{di4})
\begin{align}
\lim_{k\to\infty}kG(m,k)=\delta_{m_1+m_2,1}
\bigl[&-\theta(m_2\leq0)\theta(2k_{\Re}-a_1-a_2)+\nonumber\\
&+\theta(m_2\geq1)\theta(a_1+a_2-2k_{\Re})\bigr],\label{gr7} 
\end{align}
so that it decays with respect to $k$ at infinity and the asymptotics depends on the halfplane where $k\to\infty$. It is also easy to see that this function decays with growth of $m$:
\begin{equation}
\lim_{|m_1|+|m_2|\to\infty}G(m,k)=0.\label{gr71} 
\end{equation} 

The r.h.s.\ of (\ref{di4}) defines function that is continuous of the spectral parameter $k$ everywhere on the complex plane with exception to the points $k=a_1$ and $k=a_2$. In order to separate discontinuities at these points it is convenient to use representations (\ref{gr1}) and (\ref{gr2}):
\begin{align}
G(m,k)&=\dfrac{\delta_{m_1,0}\theta(m_2\leq0)}{a_2-a_1}+\nonumber\\
&+\dfrac{\sgn{k_{\Im}}}{2\pi{i}(a_2-a_1)m_1}\left(\left(\dfrac{\ol{k}-a_1}{k-a_1}\right)^{m_1}-1\right)
+o(1),\quad k\sim{a_1},\label{gr8} \\
G(m,k)&=-\dfrac{\theta(m_1\geq1)\delta_{m_2,0}}{a_2-a_1}-\nonumber\\
&-\dfrac{\sgn{k_{\Im}}}{2\pi{i}(a_2-a_1)m_2}\left(\left(\dfrac{\ol{k}-a_2}{k-a_2}\right)^{m_2}-1\right)
+o(1),\quad k\sim{a_2},\label{gr9} 
\end{align} 
where values at $m_1=0$ and $m_2=0$ are given by the limiting procedure under condition that $-\pi<\arg(k-a_{j})\leq\pi$. We see that limiting values of the function $G(m,k)$ at points $k=a_1,a_2$ are finite, but depend on the way on the complex plane. It is worse to mention that this property is analogous to the property of the Green's function for the heat conductivity equation on solitonic background, see, for instance \cite{ER}. We can specify the way of the approaching of the points of discontinuity, say, like
\begin{equation}
G(m,a_j)=\lim_{k_{\Re}\to{a_j}}\lim_{k_{\Im}\to0}G(m,k),\quad j=1,2,\label{gr90}
\end{equation}
so that by (\ref{gr8}) and (\ref{gr9})
\begin{equation}
G(m,a_1)=\dfrac{\delta_{m_1,0}\theta(m_2\leq0)}{a_2-a_1},\qquad G(m,a_2)=-\dfrac{\theta(m_1\geq1)\delta_{m_2,0}}{a_2-a_1}.
\label{gr91} 
\end{equation} 

The above mentioned discontinuities of the Green's function lead to singularities of its $\ol{\partial}$-derivative. Indeed, in terms of distributions we have:
\begin{align*}
\dfrac{\partial}{\partial\ol{k}}&\dfrac{1}{(k-a_2)(\zeta_2-1)-(k-a_1)(\zeta_1-1)}=\nonumber\\
&=-\dfrac{\sgn{k_{\Im}}}{2\pi{i}(\ol{k}-a_1)(\ol{k}-a_2)}
\delta\left(\zeta_1\dfrac{k-a_1}{\ol{k}-a_1}\right)\delta\left(\zeta_2\dfrac{k-a_2}{\ol{k}-a_2}\right),
\end{align*}
where $\delta(\zeta)$ denotes $\delta$-function on the unity contour,
\begin{equation}
  \delta(\zeta_j)=\sum_{n=-\infty }^{\infty }\zeta_{j}^{n},\qquad|\zeta_j|=1.\label{gr10}
\end{equation}
Thus by (\ref{di4})
\begin{equation}
\dfrac{\partial G(m,k)}{\partial\ol{k}}=-\dfrac{\sgn{k_{\Im}}}{2\pi{i}(\ol{k}-a_1)(\ol{k}-a_2)}
\Biggl(\dfrac{\ol{k}-a_1}{k-a_1}\Biggr)^{m_1}
\Biggl(\dfrac{\ol{k}-a_2}{k-a_2}\Biggr)^{m_2}.\label{gr11}
\end{equation} 

Finally, let us consider the ``integral'' form of the difference equation~(\ref{55}) on the Jost solution itself. As follows from (\ref{58:1}), it can be written in the form
\begin{equation}
\varphi(m,k)=E(m,k)+\sum_{n_1,n_2\in\Zs}\Go(m-n,k)(u_2(n)-u_1(n))\varphi(n,k),\label{di10}
\end{equation} 
where now the corresponding Green's function equals
\begin{align}
\Go(m,k)&=(k-a_1)^{m_1}(k-a_2)^{m_2}G(m,k),\label{di11}\\
\intertext{so that by (\ref{di4})}
\Go(m,k)&=\oint\limits_{|\zeta_1|=|k-a_1|}\dfrac{d\zeta_1}{2\pi i}
\oint\limits_{|\zeta_2|=|k-a_2|}\,\dfrac{d\zeta_2}{2\pi i}\dfrac{\zeta_{1}^{m_1-1}\zeta_{2}^{m_2-1}}
{\zeta_2-\zeta_1+a_2-a_1}.\label{di12} 
\intertext{This proves that $\Go(k)$ is real-valued function of $k$ even with respect to $k_{\Im}$:}
\ol{\Go(m,k)}=\Go(m,\ol{k})&=\Go(m,k),\label{di121} 
\end{align}
as follows from (\ref{gr4}), (\ref{di11}) and properties of the $G(m,k)$ given above. It is necessary to mention that if $G(m,k)$ has only discontinuities at points $k=a_1,a_2$, function $\Go(m,k)$ is singular at these points for negative $m_1$ or $m_2$. This observation shows essential difference of the discrete case from the continuous one and it is also valid for the Jost solutions themselves, see (\ref{58:1}). Because of this we work here mainly with functions $\chi(m,k)$ and $G(m,k)$ that are free from such problems.

\subsection{Properties of the Jost solutions}

Here we study properties of the Jost solution, more exactly, function $\chi(m,k)$ defined by equation~ (\ref{di3:1}), in which connection we assume below unique solvability of this equation. Because of Eqs.~(\ref{condr}) and (\ref{gr4}) reality of the potential $u(m)$ is equivalent to condition
\begin{equation}
\ol{\chi(m,k)}=\chi(m,\ol{k}),\label{di9} 
\end{equation} 
while second equality in (\ref{gr4}) shows that function
\begin{equation}
\t\chi(m,k)=\biggl(\dfrac{\ol{k}-a_1}{k-a_1}\biggr)^{m_1}
\biggl(\dfrac{\ol{k}-a_2}{k-a_2}\biggr)^{m_2}\chi(m,\ol{k}),\label{di9-1} 
\end{equation} 
obeys integral equation
\begin{align}
\t\chi(m,k)&=\biggl(\dfrac{\ol{k}-a_1}{k-a_1}\biggr)^{m_1}
\biggl(\dfrac{\ol{k}-a_2}{k-a_2}\biggr)^{m_2}+\nonumber\\
&+\sum_{n_1,n_2\in\Zs}G(m-n,k)\bigl(u_2(n)-u_1(n)\bigr)\t\chi(n,k),\label{di9-2} 
\end{align} 
i.e., equation with the same kernel as in (\ref{di3:1}). 

Asymptotic behavior of $\chi(m,k)$ follows thanks to (\ref{di3:1}), (\ref{gr7}) and (\ref{gr71}):
\begin{equation}
\lim_{k\to\infty}\chi(m,k)=1,\qquad\lim_{|m_1|+|m_2|\to\infty}\chi(m,k)=1,\label{di9-3} 
\end{equation} 
and for the second term of $1/k$ expansion we get by (\ref{gr7})
\begin{align*}
&k(\chi(m,k)-1)\to\\
&\to-\theta(2k_{\Re}-a_1-a_2)\sum_{n=m_2}^{\infty}
\bigl(u(m_1+m_2-n-1,n+1)-u(m_1+m_2-n,n)\bigr)+\\
&\quad+\theta(a_1+a_2-2k_{\Re})\sum_{n=-\infty}^{m_2-1}
\bigl(u(m_1+m_2-n-1,n+1)-u(m_1+m_2-n,n)\bigr)\bigr],
\end{align*}
that in the case under consideration of $u(m)$ rapidly decaying when $m\to\infty$ gives
\begin{equation}
\lim_{k\to\infty}k(\chi(m,k)-1)=u(m).\label{di14} 
\end{equation} 
This limiting values is independent of the halfplane of $k$ in contrast to (\ref{gr7}). It is worth to mention that from the difference equation (\ref{58:5}) we get the asymptotics behavior  in the form $k(\chi_2(m,k)-\chi_1(m,k))\to u_2(m)-u_1(m)$ only. In fact it is equivalent to (\ref{di4}) thanks to the asymptotic decaying of the potential and the second equality in (\ref{di9-3}). 

It is clear that discontinuities of the Green's function at points $k=a_1$ and $a_2$ are inherited by $\chi(m,k)$ and in analogy to (\ref{gr90}) we denote
\begin{equation}
\chi(m,a_j)=\lim_{k_{\Re}\to{a_j}}\lim_{k_{\Im}\to0}\chi(m,k),\quad j=1,2.\label{di14-1}
\end{equation} 

\subsection{Time evolution and Inverse problem}
Time evolution, i.e., dependence of $\chi(m,k)$ on $m_3$ is switched on by means of (\ref{58:6}) and for the Jost solution itself it follows by (\ref{58:1}). Let us introduce scattering data and find out their evolution. The departure from analyticity of $\chi(m,k)$ is given by the $\ol{\partial}$-differentiation of Eq.~ (\ref{di3:1}). Thanks to (\ref{gr11}) we have
\begin{align}
\dfrac{\partial\chi(m,k)}{\partial\ol{k}}&=\biggl(\dfrac{\ol{k}-a_1}{k-a_1}\biggr)^{m_1}
\biggl(\dfrac{\ol{k}-a_2}{k-a_2}\biggr)^{m_2}r(k,m_3)+\nonumber\\
&+\sum_{n_1,n_2\in\Zs}G(m,n,k)(u_2(n)-u_1(n))\dfrac{\partial\chi(n,k)}{\partial\ol{k}}.\label{di13} \\
\intertext{Here we introduced scattering data $r(k,m_3)$ defined by the equality}
r(m_3,k)&=-\dfrac{\sgn{k_{\Im}}}{2\pi{i}(\ol{k}-a_1)(\ol{k}-a_2)}\times\nonumber\\
&\times\sum_{m_1,m_2\in\Zs}\biggl(\dfrac{k-a_1}{\ol{k}-a_1}\biggr)^{m_1}
\biggl(\dfrac{k-a_2}{\ol{k}-a_2}\biggr)^{m_2}(u_2(m)-u_1(m))\chi(m,k).\label{di18} 
\end{align} 
Because of Eq.~(\ref{di9}) (i.e., because of reality of the potential $u(m)$) we have that $r(k,m_3)$ obeys
\begin{equation}
\ol{r(m_3,k)}=r(m_3,\ol{k}).\label{di19} 
\end{equation} 
Under assumption of the unique solvability of the problem (\ref{di9-2}) we get by (\ref{di13}) that  $\partial\chi(m,k)/\partial\ol{k}= r(k,m_3)\t\chi(m,k)$, or thanks to (\ref{di9-1}) that
\begin{equation}
\dfrac{\partial\chi(m,k)}{\partial\ol{k}}=\biggl(\dfrac{\ol{k}-a_1}{k-a_1}\biggr)^{m_1}
\biggl(\dfrac{\ol{k}-a_2}{k-a_2}\biggr)^{m_2}r(m_3,k)\chi(m,\ol{k}).\label{di20} 
\end{equation} 

Time evolution of the spectral data, i.e., dependence on $m_3$ trivially follows from $\ol{\partial}$-differentiation of the second equation of the Lax pair, Eq.~(\ref{58:6}), and (\ref{di13}):
\begin{equation}
r(m_3,k)=\biggl(\dfrac{\ol{k}-a_3}{k-a_3}\biggr)^{m_3}r(k),\label{di21} 
\end{equation} 
where function $r(k)$ is independent of $m_3$ and by (\ref{di18}) is uniquely defined by the initial data (\ref{CPu}). Eq.~(\ref{di18}) shows that the spectral data have integrable singularities at points $k=a_1,a_2$.

Summarizing, the inverse problem to determine $\chi(m,k)$ is given by the equation
\begin{equation}
\dfrac{\partial\chi(m,k)}{\partial\ol{k}}=R(m,k)\chi(m,\ol{k}),\label{di17} 
\end{equation} 
with normalization condition (\ref{58:3}). Here we denoted
\begin{equation}
R(m,k)=\biggl(\dfrac{\ol{k}-a_1}{k-a_1}\biggr)^{m_1}
\biggl(\dfrac{\ol{k}-a_2}{k-a_2}\biggr)^{m_2}\biggl(\dfrac{\ol{k}-a_3}{k-a_3}\biggr)^{m_3}r(k),\quad
k\in\Cs.\label{di22} 
\end{equation} 
For any $r(k)$ this function obeys (\ref{lB}), i.e., the linearized version of the Hirota difference equation (\ref{Hd}).  We also mention that because of (\ref{di21}) 
\begin{equation}
|R(m,k)|=|r(k,m_3)|=|r(k)|,\label{di221}
\end{equation} 
i.e., it is independent of $m_3$.

\subsection{Integrals of motion}
Let us introduce function
\begin{equation}
\rho(k)=\sum_{n_1,n_2\in\Zs}(u_2(m)-u_1(m))\chi(m,k).\label{im1} 
\end{equation} 
Thanks to the asymptotic decaying of the potential $u(m)$ and boundedness of the function $\chi(m,k)$ by $m$ this series converge and function  $\rho(k)$ decays when $k\to\infty$. It obeys conjugation property
\begin{equation}
\overline{\rho(k)}=\rho(\overline{k}),\label{im2} 
\end{equation} 
thanks to reality of the potential. For the $\ol{\partial}$-derivative of this function we get by (\ref{di18})--(\ref{di22})
\begin{align*}
\dfrac{\partial\rho(k)}{\partial\overline{k}}&=r(k)\sum_{n_1,n_2\in\Zs}\Bigl(\dfrac{\overline{k}-a_1}{k-a_1}\Bigr)^{m_1}
\Bigl(\dfrac{\overline{k}-a_2}{k-a_2}\Bigr)^{m_2}\bigl(u_2(m)-u_1(m))\bigr)\chi(m,\overline{k})=\nonumber\\
&=-4\pi{i}(k-a_1)(k-a_2)\sgn\bigl((a_1-a_2)k_{\Im}\bigr)|r(k)|^{2}.
\end{align*} 
Now taking (\ref{58:3}) into account we get that in terms of the scattering data function $\rho(k)$ is given by equality
\begin{equation}
\rho(k)=-4i\int{d^{2}k'}\dfrac{(k'-a_1)(k'-a_2)}{k-k'}\sgn\bigl((a_1-a_2)k'_{\Im}\bigr)|r(k')|^{2},\label{im4} 
\end{equation} 
where $dk^{2}=dk_{\Re}dk_{\Im}$. Thanks to Eq.~(\ref{im4}) this proves that  $\rho(k)$ is independent of time $m_3$ and it is the generating function of the infinite set of integrals of motion. Thus thanks to (\ref{di14}) the first nontrivial integral (the first coefficient of $1/k$ expansion) is
\begin{align}
\rho_{1}&=\sum_{n_1,n_2\in\Zs}(u_2(m)-u_1(m))u(m)=\nonumber\\
&=-4i\int{d^{2}k'}(k'-a_1)(k'-a_2)\sgn\bigl((a_1-a_2)k'_{\Im}\bigr)|r(k')|^{2}.\label{im3}
\end{align} 

\section{Darboux transformation}
\subsection{Darboux transformation and time evolution}
In the case of discrete systems time evolution is a special case of the Darboux transformation. We start with the Darboux transformation of the continuous spectrum, i.e., transformation that gives new potential $\h{u}$ and the corresponding Jost solution $\h\chi(m,k)$ that is determined by the same normalization (\ref{58:3}) and $\ol\partial$-equation (\ref{di17}) with substitution
\begin{equation}
R(m,k)\to \h{R}(m,k)=\dfrac{\ol{k}-a}{k-a}\,R(m,k)\label{dt0} 
\end{equation} 
where an arbitrary parameter $a$ must be chosen real in order to preserve property (\ref{di19}). Then by (\ref{di17}) we get for the $\ol\partial$-derivative 
\begin{align*}
&\dfrac{\partial}{\partial\ol{k}}\bigl[(k-a)\h\chi(m,k)-(k-a_j)\chi^{(j)}(m,k)\bigr]=\nonumber\\
&\qquad=R(m,k)\bigl[\ol{(k-a)\h\chi(m,k)-(k-a_j)\chi^{(j)}(m,k)}\bigr],\quad j=1,2,3,
\end{align*} 
where difference in brackets is bounded at $k\to\infty$. Indeed, in this limit
\begin{equation}
(k-a)\h\chi(m,k)-(k-a_1)\chi^{(j)}(m,k)\to \h{u}(m)-u^{(j)}(m)+a_1-a,\label{dt01} 
\end{equation} 
while $\h{u}$ is given by asymptotics of the Jost solution $\h\chi(m,k)$ like in (\ref{di14}). Thus under assumption of the unique solvability of the Inverse problem (\ref{58:3}) and (\ref{di17}) we derive for the new Jost solution equation
\begin{equation}
(k-a)[\h\chi(m,k)-\chi(m,k)]=(k-a_j)\chi_j(m,k)-(\h{u}(m)-u^{(j)}(m)\chi(m,k).\label{dt7} 
\end{equation} 

It is clear that in the limit $k\to{a}$ the Jost solution $\h\chi$ drops out, so by means of this transformation the potential $\h{u}$ is given by means of
\begin{equation}
\h{u}(m)=u^{(j)}(m)-(a-a_j)\dfrac{\chi_{j}(m,a)}{\chi(m,a)},\quad j=1,2,3,\label{dt6} 
\end{equation} 
in other words, is given explicitly in terms of the original potential and value of its Jost solution at the value of the spectral parameter equal to the parameter $a$ of the Darboux transformation (\ref{dt0}). Potential $\h{u}$ can be excluded from (\ref{dt7}) and we get
\begin{equation}
\h\chi(m,k)=\dfrac{k-a_j}{k-a}\chi^{(j)}(m,k)-\dfrac{(a-a_j)\chi^{(j)}(m,a)}{\chi(m,a)}\chi(m,k).\label{dt8} 
\end{equation} 
Setting here $k=a_j$ (we use definition (\ref{di14-1}) for the values of function  $\chi(m,k)$ at points $a_j$) we derive
\begin{equation}
\dfrac{\h\chi(m,a_j)}{\chi(m,a_j)}=\dfrac{\chi^{(j)}(m,a)}{\chi(m,a)},\label{dt9} 
\end{equation} 
so that (\ref{dt8}) can be written in the symmetric form:
\begin{equation}
\dfrac{1}{k-a_j}\left[\dfrac{\h\chi(m,k)}{\chi(m,k)}-\dfrac{\h\chi(m,a_j)}{\chi(m,a_j)}\right]=
\dfrac{1}{k-a}\left[\dfrac{\chi^{(j)}(m,k)}{\chi(m,k)}-\dfrac{\chi^{(j)}(m,a)}{\chi(m,a)}\right].\label{dt10} 
\end{equation} 

Let us choose $a=a_3$. Then in terms of the function $v(m)$ (see (\ref{vu})) we get from above:
\begin{align}
&v^{(3)}(m)=v^{(j)}(m)-(a_3-a_j)\dfrac{\chi^{(j)}(m,a_3)}{\chi(m,a_3)},\label{dt11} \\
&\chi^{(3)}(m,k)=\dfrac{k-a_j}{k-a_3}\,\chi^{(j)}(m,k)-\dfrac{(a_3-a_j)\chi^{(j)}(m,a_3)}{(k-a_3)\chi(m,a_3)}\,\chi(m,k),\label{dt12} 
\end{align} 
where $j=3$ gives identity. We notice that by (\ref{di22}) it is clear that transformation of the scattering data in (\ref{dt0}) in case of $a=a_3$ is nothing but shift of the time variable $m_3\to{m_3}+1$. Correspondingly, relations (\ref{dt11}) and (\ref{dt12}) at $j=1$ or $j=2$ demonstrate this property in terms of the potential and Jost solution. On the other side, performing consequent Darboux transformations with values of parameter $a$, say, $a_4$, $a_5$, etc., different from the original $a_1,a_2,a_3$, we introduce dependence on the new, ``highest'' times $m_4$, $m_5$, etc. Taking (\ref{di22}) into account it is clear that in respect to any three ``times" $m_i$, $m_j$ and $m_k$ ($i\neq{j}\neq{k}$) we get the same Hirota difference equation. 

\subsection{$\tau$-function}
Here we present relation of the IST approach developed above with the standard for the discrete systems $\tau$-function formulation. Taking ill definiteness of the Jost solution at points $k=a_1,a_2$ into account we fix values at these points, say, as in (\ref{di14-1}). We write  (\ref{58:5}) as equation on the function $v(m)$, see (\ref{vu}), using notation of the kind (\ref{upper}) and (\ref{fd}):
\begin{equation*}
v^{(2)}-v^{(1)}=\dfrac{(k-a_{2})\chi^{(2)}(k)}{\chi(k)}-\dfrac{(k-a_{1})\chi^{(1)}(k)}{\chi(k)},
\end{equation*}
omitting everywhere dependence on $m$. Setting here $k=a_1$ and $k=a_2$ we get two equalities:
\begin{align}
&v^{(2)}-v^{(1)}=(a_1-a_2)\dfrac{\chi^{(1)}(a_2)}{\chi(a_2)}=(a_1-a_2)\dfrac{\chi^{(2)}(a_1)}{\chi(a_1)},\label{t1}\\ 
\intertext{and in the same way by (\ref{60:1}) and (\ref{58:6}) correspondingly}
&v^{(3)}-v^{(2)}=(a_2-a_3)\dfrac{\chi^{(2)}(a_3)}{\chi(a_3)}=(a_2-a_3)\dfrac{\chi^{(3)}(a_2)}{\chi(a_2)},\label{t2}\\ 
&v^{(1)}-v^{(3)}=(a_3-a_1)\dfrac{\chi^{(3)}(a_1)}{\chi(a_1)}=(a_3-a_1)\dfrac{\chi^{(1)}(a_3)}{\chi(a_3)},\label{t3} 
\end{align}
that results also from the symmetry under cyclic permutation of indexes. This means that
\begin{equation}
\dfrac{\chi^{(i)}(a_j)}{\chi(a_j)}=\dfrac{\chi^{(j)}(a_i)}{\chi(a_i)},\quad i,j=1,2,3.\label{t4} 
\end{equation} 
Let us introduce $\tau$-function by means of the system of equations:
\begin{align}
&\tau^{(1)}(m)=(a_3-a_1)^{m_3}(a_1-a_2)^{m_2}\chi(m,a_1)\tau(m),\label{t5}\\ 
&\tau^{(2)}(m)=(a_1-a_2)^{m_1}(a_2-a_3)^{m_3}\chi(m,a_2)\tau(m),\label{t6}\\ 
&\tau^{(3)}(m)=(a_2-a_3)^{m_2}(a_3-a_1)^{m_1}\chi(m,a_3)\tau(m),\label{t7} 
\end{align}
that are compatible thanks to (\ref{t4}). Substitution of $\chi(m,a_j)$ from these relations into (\ref{t1})--(\ref{t3}) gives (\ref{vt1})--(\ref{vt3}). On the other side, summing up (\ref{t1})--(\ref{t3}) we get
\begin{equation}
(a_1-a_2)\dfrac{\chi^{(2)}(a_1)}{\chi(a_1)}+\text{cycle}(1,2,3)=0,\label{t8} 
\end{equation} 
that thanks to (\ref{t5})--(\ref{t7}) is (\ref{Ht}).

\section{Soliton solutions}

Soliton solutions for the Hirota difference equation are well known in the literature (see \cite{Hirota1}--\cite{Miwa}).  It is also known, see \cite{Gramm}, that this equation admits also lump solutions. Here we introduce soliton solutions by means of the following construction. Let we have two numbers: 
\begin{equation}
N_a,N_b\geq1,\label{s1} 
\end{equation} 
and set of $N$ real parameters $\varkappa_n$ that we can choose to be ordered:  $\varkappa_1<\varkappa_2<\cdots<\varkappa_N$, where
\begin{equation}
N=N_a+N_b.\label{s2} 
\end{equation} 
Let $\chi(m,k)$ be a meromorphic function of $k$ that has poles at points $k=\varkappa_{n_1},\ldots,\varkappa_{n_{N_b}}$, where $\{n_1,\ldots,n_{N_b}\}$ is a subset of $\{1,\ldots,N\}$. Let us rescale the Jost solution:
\begin{equation}
\chi(m,k)\to\chi(m,k)\prod_{j=1}^{N_b}(k-\varkappa_{n_j})^{-1},\label{s2:1} 
\end{equation} 
so that the new one is a polynomial of order $k^{N_b}$ with the unity coefficient at higher power. Thanks to (\ref{di14}) we have
\begin{equation}
\dfrac{\chi(m,k)}{k^{N_b}}=1+\dfrac{1}{k}\Bigl(u(m)-\sum_{j=1}^{N_b}\varkappa_{n_j}\Bigr).\label{s2:2} 
\end{equation} 
 Thus
\begin{equation}
\chi(m,k)=k^{N_b}+\sum_{l=1}^{N_b}k^{l-1}X(l,m),\label{s3} 
\end{equation} 
where $X(l,m)$ are some coefficients to be determined. For this aim we use (\ref{58:1}) with $\chi(x,k)$ substituted from the latter equality and on values of the Jost solution at points $k=\varkappa_{n}$  impose $N_b$ conditions:
\begin{equation}
\bigl(\varphi(m,\varkappa_1),\ldots,\varphi(m,\varkappa_N)\bigr)D=0.\label{s4} 
\end{equation} 
where $D$ is $N\times{N_b}$-matrix with at least two nonzero maximal minors. This condition gives linear system of equations to determine uniquely $X(l,m)$. To describe solution of this system we use here the following notation: let $V$ be incomplete Vandermond $(N_{b}+1)\times{N}$-matrix
\begin{equation}
V=\left(\begin{array}{lll}
1 & \ldots & 1 \\
\vdots &  & \vdots \\
\varkappa_{1}^{N_{b}} & \ldots & \varkappa_{N}^{N_{b}}
\end{array}\right) ,\label{s9}
\end{equation} 
and let $V(l)$ is matrix $V$ with removed $l$-th row (i.e., $N_{b}\times{N}$-matrix). We also need two diagonal $N\times{N}$-matrices:
\begin{align}
&E(m)=\diag\{E(m,\varkappa_1),\ldots,E(m,\varkappa_N)\}\label{s11}\\
\intertext{see (\ref{58:2}), and}
&k-\varkappa=\diag\{k-\varkappa_1,\ldots,k-\varkappa_N\}.\label{s10}
\end{align} 
Let also $Y(l,m)$ denote determinant of $N_b\times{N_b}$-matrix
\begin{equation}
Y(l,m)=(-1)^{N_b+1-l}\det(V(l)E(m)D),\qquad Y(m)=Y(N_{b}+1,m).\label{s4:1} 
\end{equation} 
Then it is easy to see that
\begin{equation}
X(l,m)=-\dfrac{Y(l,m)}{Y(m)},\label{s4:0} 
\end{equation} 
Now by (\ref{s3}) we readily  get
\begin{equation}
\chi(m,k)=\dfrac{Z(m,k)}{Y(m)},\label{s6} 
\end{equation}
where 
\begin{equation}
Z(m,k)=\det\bigl(V(N_b+1)(k-\varkappa)E(m)D\bigr),\label{s7}
\end{equation}
where notation (\ref{s10}) was used. Thanks to (\ref{s2:2}) we get
\begin{equation}
u(m)=\sum_{j=1}^{N_b}\varkappa_{n_j}-\dfrac{Y(N_b,m)}{Y(m)}.\label{s12} 
\end{equation} 
Let us construct the corresponding $\tau$-function. Thanks to (\ref{58:2}) it is easy to see that
\begin{equation}
\chi(m,a_n)=(-1)^{N_b}\dfrac{Y^{(n)}(m)}{Y(m)}.\label{s13} 
\end{equation} 
thus, using definitions (\ref{t5})--(\ref{t7}) we see that $\tau$-function equals (up to a some constant factor)
\begin{equation}
\tau(m)=(-1)^{N_b(m_1+m_2+m_3)}(a_3-a_1)^{m_1m_3}(a_1-a_2)^{m_2m_1}(a_2-a_3)^{m_3m_2}Y(m).\label{s14} 
\end{equation} 

As an example of this generic construction we present one-soliton solution that equals
\begin{align}
&u(m)=\dfrac{\varkappa_2-\varkappa_1}{1+cf(m)},\label{1s1} \\
\intertext{where $c$ a real constant and}
&f(m)=\dfrac{E(m,\varkappa_2)}{E(m,\varkappa_1)}\equiv
\biggl(\dfrac{\varkappa_2-a_1}{\varkappa_1-a_1}\biggr)^{m_1}
\biggl(\dfrac{\varkappa_2-a_2}{\varkappa_1-a_2}\biggr)^{m_2}
\biggl(\dfrac{\varkappa_2-a_3}{\varkappa_1-a_3}\biggr)^{m_3}.\label{1s2} 
\end{align}
Already this example shows that the consideration here was formal in the sense that denominator in (\ref{s12}) (i.e., $\tau$-function) can take zero values, so solution can be singular for some values of $m$. This situation reminds analogous problem well known in the case of KPII equation. But there independent variables are continuous and absence of singularities is equivalent to the sign definiteness of  the $\tau$-function. Here when the independent variables run through discrete values, it is obvious that situation is more involved: $\tau$-function can change sign without passing through zero. It is also necessary to mention that (\ref{s12}), in fact, defines solution up to a constant and as it is easy to see generically this solution has different constant limits as depending on the direction $m_1$ and $m_2$ tends to infinity. So, strictly speaking soliton solutions do not fit in the class of solutions for which the IST was developed in the previous sections. Soliton solutions interpolate between different constants on the $m$-infinity and one has to develop version of the IST that enables consideration of such solutions. Another property, specific for the soliton solutions of the Hirota difference equation is existence of a resonant solitons, i.e., solitons where parameters $\varkappa_{i}$ coincide with some of parameters $a_1,a_2,a_3$. One soliton solution (\ref{1s1}) shows that in the corresponding limit solution exists, but its properties can be rather strange. This special class of solitons deserves to be studied in detail.

\textbf{Acknowledgement.} This work is supported in part by the Russian Foundation for Basic Research (grants \# 13-01-12405 and \# 14-01-00860) and by the Program of RAS ``Fundamental Problems of the Nonlinear Dynamics''.

\end{document}